\documentclass[10pt,journal,compsoc]{IEEEtran}
\usepackage{amsmath,amssymb,amsfonts}
\usepackage{algorithmic}
\usepackage{graphicx}
\usepackage{textcomp}
\usepackage{color}
\usepackage{amsmath}
\usepackage{multirow}
\usepackage{verbatim}
\usepackage{float}
\ifCLASSOPTIONcompsoc
  \usepackage[nocompress]{cite}
\else
  \usepackage{cite}
\fi

\ifCLASSINFOpdf
 
\else
 
\fi

\begin{document}

\title{Segmentation Network with Compound Loss Function for Hydatidiform Mole Hydrops Lesion Recognition}

\author{Chengze Zhu, Pingge Hu, Xianxu Zeng, Xingtong Wang, Zehua Ji and Li Shi

\IEEEcompsocitemizethanks{\IEEEcompsocthanksitem C. Zhu, P. Hu, X. Wang, Z. Ji and L. Shi are with the Department of Automation, Tsinghua University, Beijing, 100084 P.R.China. E-mail:zhucz18@tsinghua.org.cn; hpg18@mails.tsinghua.edu.cn; xingtong21@mails.tsinghua.edu.cn; jizh18@mails.tsinghua.edu.cn; shilits@mail.tsinghua.edu.cn
\IEEEcompsocthanksitem X. Zeng is with the Department of Pathology, the Third Affiliated Hospital of Zhengzhou University,
Zhengzhou, Henan, 450052 P.R.China. 
E-mail: xianxu77@163.com).
\IEEEcompsocthanksitem C. Zhu and P. Hu are with equal contribution.

}% <-this % stops an unwanted space

}

\markboth{}%
{Shell \MakeLowercase{\textit{et al.}}: Bare Demo of IEEEtran.cls for Computer Society Journals}

\IEEEtitleabstractindextext{%
\begin{abstract}
Pathological morphology diagnosis is the standard diagnosis method of hydatidiform mole. As a disease with malignant potential, the hydatidiform mole section of hydrops lesions is an important basis for diagnosis. Due to incomplete lesion development, early hydatidiform mole is difficult to distinguish, resulting in a low accuracy of clinical diagnosis. As a remarkable machine learning technology, image semantic segmentation networks have been used in many medical image recognition tasks. We developed a hydatidiform mole hydrops lesion segmentation model based on a novel loss function and training method. The model consists of different networks that segment the section image at the pixel and lesion levels. Our compound loss function assign weights to the segmentation results of the two levels to
calculate the loss. We then propose a stagewise training method to combine the advantages of various loss functions at different levels. We evaluate our method on a hydatidiform mole hydrops dataset. Experiments show that the proposed model with our loss function and training method has good recognition performance under different segmentation metrics. 
\end{abstract}

}

\maketitle

\IEEEdisplaynontitleabstractindextext

% \fi
\IEEEpeerreviewmaketitle

\IEEEraisesectionheading{\section{Introduction}\label{sec:introduction}}

\label{sec:introduction}
\IEEEPARstart{H}{ydatidiform} mole(HM) is one of the most common gestational trophoblastic diseases(GTD), which occur in about 1 in 500-1000 pregnancies[1]-[2]. Since there is a certain probability that HM will develop into invasive HM and choriocarcinoma, most HM fetuses are unviable, or HM grows into a teratoma [3]-[6]. HM is commonly found in women under age 17 or over age 35, and can be partial or complete. Although moles can be identified using different types of diagnostic methods, the final diagnosis must be confirmed by pathologists. Pathological diagnosis of tissue section is the gold standard for the diagnosis of HM. Pathologists generally use microscopes of $5 \times 10$ times and $10\times10$ times  magnification to observe multiple sections of subjects and make comprehensive diagnoses according to experience and the morphology of the sections. Fig. 1  shows a well-developed complete HM under a microscope. Pathologists mainly diagnose by observing the villi characteristics of HM in the sections [5]. With partial HM, there is local hydrops of villi stroma and local  trophoblast hyperplasia at the edge of the villi, and with complete HM, there is entire hydrops of villi stroma and diffuse hyperplasia of trophoblastic cells at the edge of the villi [6]. Therefore, hydrops lesions are an important basis for HM diagnosis.

In actual pathological work, HM before 12 weeks of pregnancy is often morphologically confused with non-HM pregnancy and other diseases due to incomplete development [7]-[8], pathologists must spend substantial time on the diagnosis, resulting in low detection efficiency. There is a need for an auxiliary diagnosis system for HM that can improve the diagnostic accuracy to reduce missed diagnoses and misdiagnoses. 

Computer-aided diagnosis has been widely adopted in clinical practice recent years, especially the neural network based medical imaging recognition algorithms[28]-[37]. An image semantic segmentation network based on deep learning has been proved to effectively increase the accuracy and efficiency of diagnosis in many medical imaging lesion segmentation tasks. However, there is rare research on intelligent diagnosis of HM lesions based on deep learning[3]. To address the above issues, we propose an intelligent auxiliary diagnosis method that can identify HM lesions under a microscope in real time. The contributions of this paper are as follows: (1) A semantic segmentation model for HM hydrops lesion segmentation is constructed. HM hydrops lesion segmentation models under different networks, feature-extraction networks, and loss functions are tested and evaluated. Experimental results confirm that the network model has a good segmentation effect on hydrops datasets. (2) A compound loss function combined with pixel- and lesion-level loss for multiple evaluation metrics is constructed that, compared with the traditional loss function, shows good performance on pixel- and lesion-level evaluation metrics. We also propose a stagewise training method of multiple loss functions, which we experimentally show significantly improves segmentation results. 

Other than assisting diagnosis of HM, this model can also be generalized since its components are not dedicated to only recognize the HM lesion, but designed to extract the features we need for the corresponding diagnosing process. Therefore, our proposed method is also able to assist diagnosing other diseases by extracting the corresponding pathological features. 

\begin{figure}[h]
	\centering
	\includegraphics[width=6cm]{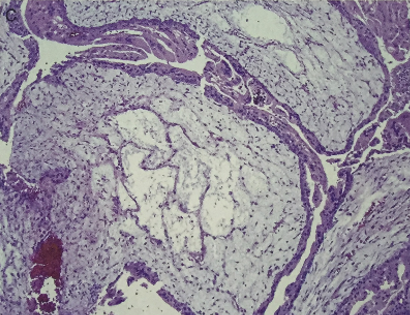}
	\caption{Complete HM under microscope.}
	\label{fig1}
\end{figure}

\section{Related Work}
Here we discuss semantic segmentation artificial neural network, which has been widely applied to medical images, and pathological section image segmentation.

\subsection{Semantic Segmentation Network}
Semantic segmentation is pixel-level classification. Before the advent of deep learning, the traditional classifier is generally designed for a single category and is greatly limited by features [9]-[13]. Deep learning has simplified the pipeline of semantic segmentation and obtained better segmentation results. The fully convolutional network (FCN) [14] extracts features and samples high-level semantic features to specified dimensions to obtain the final prediction results, which naturally form an encoder-decoder framework. Classical algorithms, such as U-Net [15], the Feature Pyramid Network  (FPN) [16], SegNet [17], PSPNet [18], and LinkNet [19], all have an encoder-decoder structure. The DeepLab series introduces an atrous convolution method, which controls the receptive field size of a model and can therefore obtain feature information in different ranges [20]-[22]. DeepLabv3+ [22] combines spatial pyramid pooling and an encoder-decoder structure, and also exploits a more powerful network by using modified aligned exception and atrous separable revolution.

\subsection{Pathological Section Image Segmentation}
Most early lesion identification methods for pathological section are unsupervised and semi-supervised. These methods use lesion’s own feature in the section images as descriptors, uses threshold, cluster, similarity measuring and other methods to segment the lesion.[23]-[27] For examples, the machine learning approach employs fuzzy C-Means clustering with hue, saturation and value color space [27] in order to divide the hydatidiform mole villi area correctly. Wavelet based texture features [23] were used to segment lesions in hyperspectral human colon tissue cell images. MLSeg [25] defines a new set of high-level texture descriptors to represent prior knowledge in colon tissue and uses it in unsupervised multi-level segmentation algorithms. Semi-supervised learning for both spectral dimension reduction and hierarchical pixel clustering [24] were used for hyperspectral images of tissue samples and successfully segmented the pixels of different cell types in the images. Although the traditional method of lesion recognition has been successfully applied in pathological practice, the traditional algorithm is limited by the weak expression ability of feature operator, which has the loss of recognition accuracy and generalization performance, just like the disadvantages of manual feature in image processing.

Image segmentation network based on deep learning has strong learning ability and greatly improves the accuracy of lesion recognition in pathological images. Segmentation networks commonly used for assisiting diagnosis include CNN [31]-[33], U-Net [15][29][30], DeepLab Series [34], etc. On the basis of the above network, combined with the characteristics of pathological image data set, a more targeted network model is proposed. DRA-Net combined with the new computer-aided cancer diagnosis framework based on Session Histopathological Image Recommendation (SHIR) [28] successfully learned the pathological knowledge of lesion recognition by using only WSI tags. A human-like automatic diagnostic network [35] designed a structure including a scanning network (s-net), a diagnostic network (d-net), and an aggregation network (a-net) for urothelial carcinoma biopsy data of bladder cancer subjects.

Neural network already has preliminary results in pathologic diagnosis of hydatidiform mole. P. Pal et al. [3] classified pathological sections of hydatidiform mole villi into normal, PHM, or CHM categories based on various characteristics of hydatidiform mole using three fully connected networks. The overall accuracy of the validation dataset is able to reach 86.1\%. However, this method is based on the hydatidiform mole section image with complete structures and edges, can only extract the superficial features of the image and classify the whole image. The algorithm has a simple structure and does not have the capability to locate and segment the lesion from the whole image. In this paper, our method uses an efficient semantic segmentation network to accurately segment the input pathological images and provide doctors with more convincing results.

\section{Proposed Method}
\subsection{Dataset and Labeling}
The data used in this paper are from \emph{the Third Affiliated Hospital of Zhengzhou University}. The data usage was approved by \emph{the Third Affiliated Hospital of Zhengzhou University Ethic Committee}. We selected and scanned 157 HM sections from 59 subjects with a \emph{Motic} section scanner as the main section dataset. Subjects’ ages ranged from 25 to 38, and the amenorrhea time ranged from 30 to 92 days. Most sections had complete or partial HM, and the remaining sections had several diseases that are easily confused with HM.

With the help of pathologists, we performed hydrops, hyperplasia, and villi labeling on the scanned section images. All labeling results were examined and approved by clinical pathologists. We used \emph{Motic} digital section assistant system to obtain section images and annotation files. After data conversion and image processing, we obtained the HM section and HM section hydrops lesion label images shown in Fig. 2.

\begin{figure}[!t]
	\centerline{\includegraphics[width=\columnwidth]{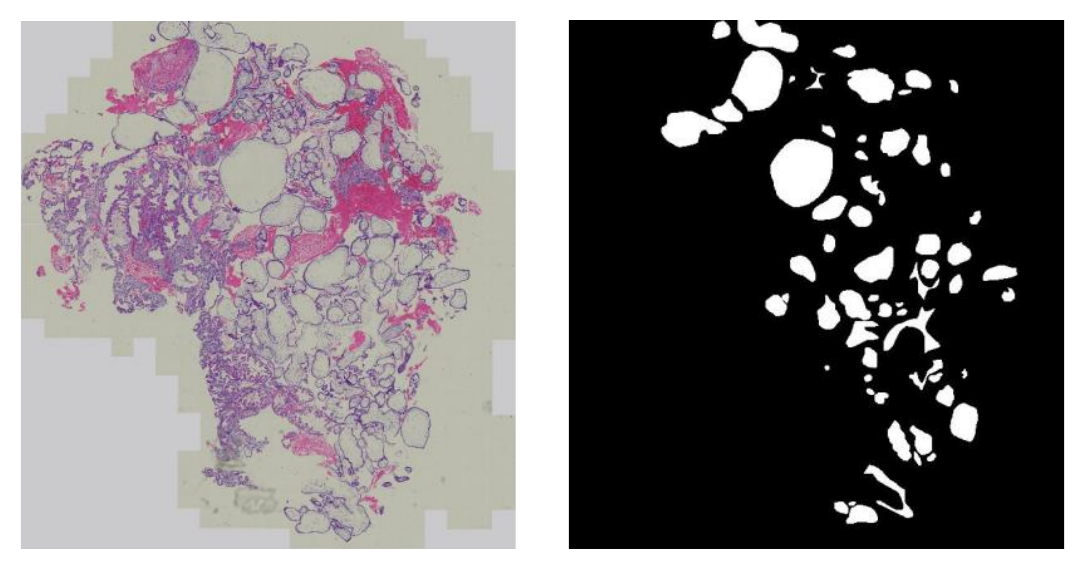}}
	\caption{Scan and label mask of HM. Left: HM section scan under microscope; Right: Label mask of HM.}
	\label{fig2}
\end{figure}

\subsection{Diagnostic Deep Network}
Mainstream semantic segmentation network structures include DeepLabv3+, U-Net, and FPN, all of which have an encoder-decoder architecture on a macro level. We use encoder-decoder architecture to construct the hydrops lesion segmentation model. As shown in Fig. 3(a), the input is the preprocessed HM section, and the output is the label map of the hydrops lesion of the HM section.

In the decoder in Fig. 3(a), convolution and upsampling ensure that the feature image remains the same size, and P1, P2, and P3 at different scales obtained by the feature-extraction network are merged. Networks such as DeepLabv3+, UNet, and FPN use different merge methods.

\begin{figure*}[!t]
	\centerline{\includegraphics[width=18.1864cm]{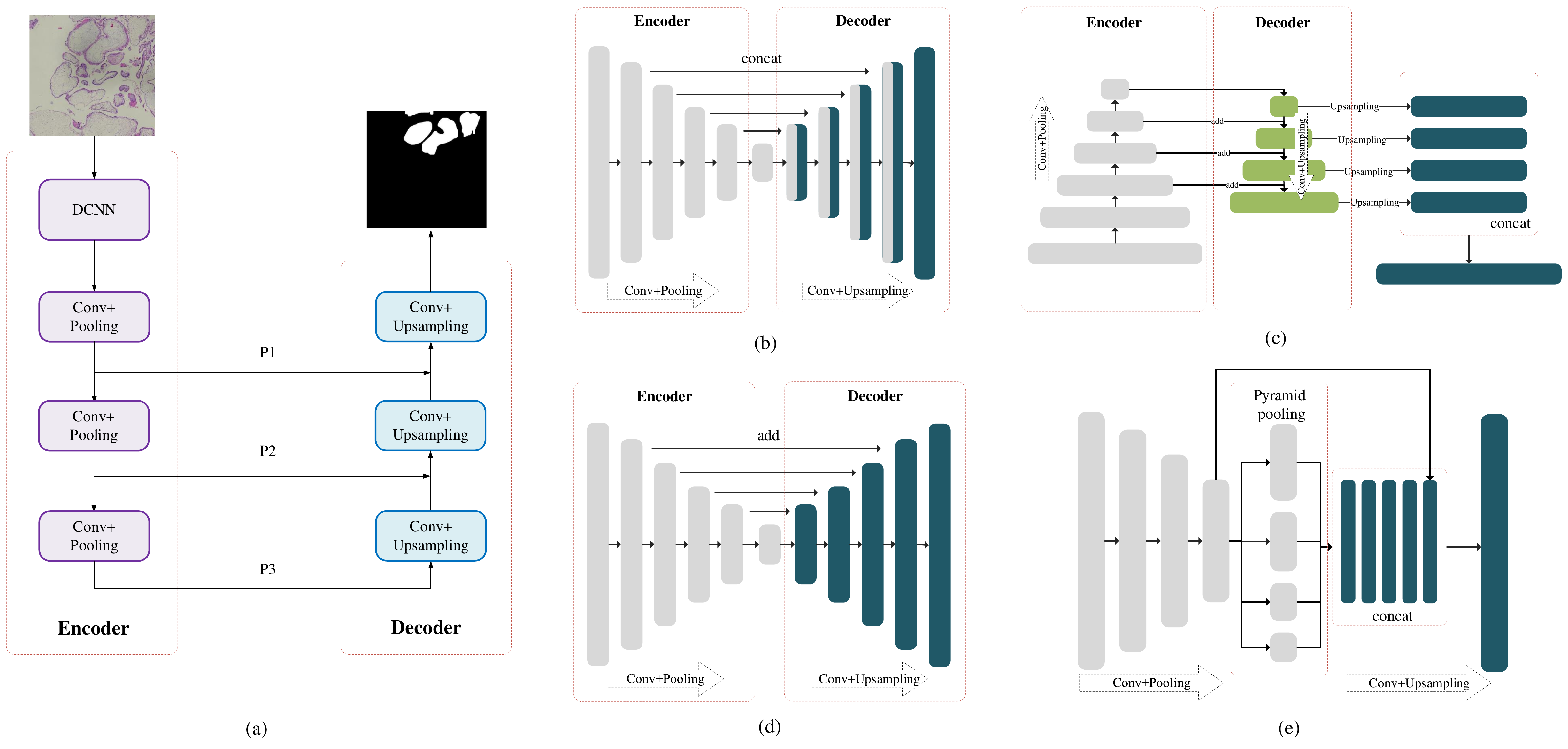}}
	\caption{Overview of the hydrops lesion segmentation model.(a) Framework of hydrops lesion segmentation model. (b) U-Net. (c) FPN. (d) LinkNet. (e) PSPNet.}
	\label{fig}
\end{figure*}

U-Net is widely used in the semantic segmentation task of medical image lesion segmentation. Its structure is shown in Fig. 3(b). The encoder extracts multiscale features, whereas the output features in decoder engages the same size corresponding to the encoder. Then, the feature maps of the encoder and decoder are combined (concat). U-Net overcomes insufficient upsampling information by merging feature images of the same resolution between the encoder and decoder and provides multiscale information for semantic segmentation by retaining deep and shallow feature information.

Unlike U-Net, FPN adds two images directly to fuse the image, and the fused feature map is upsampled to obtain the multiscale prediction results. The image segmentation results obtained from the upsampling of each layer are combined to obtain the results.

The pyramid scene parsing network (PSPNet) constructs a pyramid pooling module to obtain multiscale information. As shown in Fig. 3(e), PSPNet inputs the feature map to the pyramid pooling module. The pooling feature map takes convolutions and upsamples to obtain the same scale as the input feature image of the pyramid pooling layer. Four feature maps, with different scales and input feature maps of the pyramid pooling layer, are merged. LinkNet has a structure similar to U-Net, but the different scales output by the encoder are directly added to the corresponding characteristic images of the decoder, which is similar to FPN. LinkNet also optimizes the network structure, meaning it obtains better semantic segmentation results without adding parameters (see Fig. 3(d)).

\begin{figure}[!t]
	\centerline{\includegraphics[width=\columnwidth]{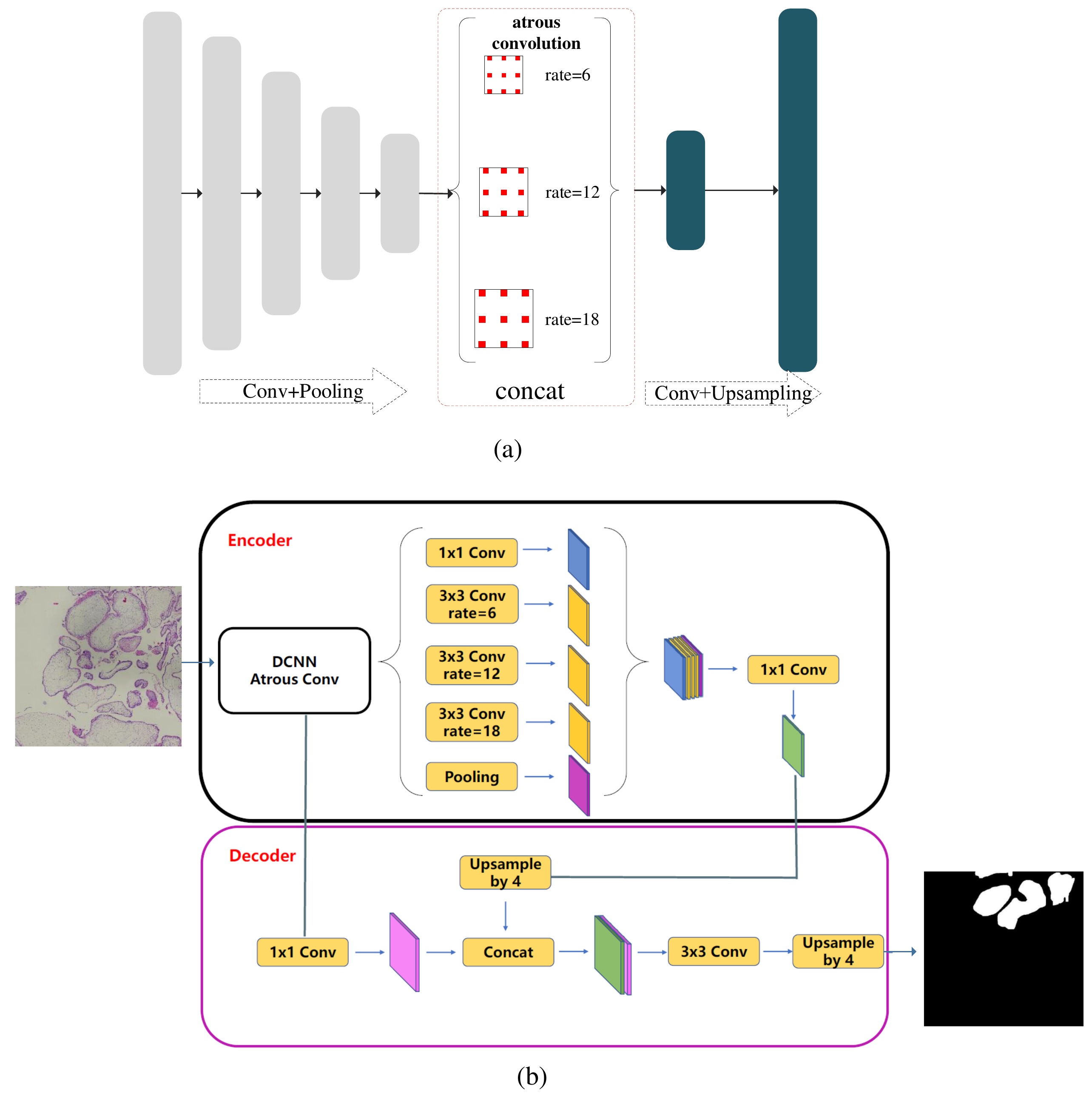}}
	\caption{ (a) Structure of DeepLabv3+. (b) DeepLabv3+ network structure on HM lesion segmentation.}
	\label{fig}
\end{figure}

DeepLabv3+ is the latest version of DeepLab, the main network used in this paper. The network structure is shown in Fig. 4. DeepLabv3+ uses atrous convolution, which does not apply to the adjacent $3\times3$ feature images but to the nine feature points of the $3\times3$ interval rate. Feature convolution at different scales is completed with different rate. In particular, when the rate is 1, atrous convolution is the same as traditional convolution. Atrous convolution can expand the receptive field and capture multiscale feature information, as shown in Fig. 4(a). DeepLab parallels the atrous convolutions of multiple scales and combines the output results. Since the rate can be freely selected, it can adapt to different scales of lesion segmentation during network parameter adjustment, which is of great significance for the lesion segmentation of HM hydrops.

\subsection{Compound Loss Function}
We conducted experiments on commonly used loss functions BCELoss, DiceLoss, IoULoss, and IoULoss-based FocalLoss. Although it can be seen that IoULoss and BCELoss both have good performance on the dataset, they are insensitive to lesion-level evaluation metrics. Detailed experiments and results are shown in the Experiment sections. Also, in clinical practice, doctors focus on different sections, so the model will require a higher recall rate. Therefore, we propose a compound loss function that not only takes into account pixel- and lesion-level loss but also take both recall loss and precision loss into consideration: 
\begin{equation}
\begin{aligned}
CompoundLoss&=F_{focal}(\alpha \times LesionLoss+(1-\alpha)\\
&\quad\times PixelLoss)\vspace{1ex}
\end{aligned}
\label{eq}
\end{equation}
\noindent
where
\begin{equation}
\begin{aligned}
F_{focal}(x)&=-k\times(1-x)^{\gamma}\times log(x)\vspace{1ex}
\end{aligned}
\label{eq}
\end{equation}
\noindent
where $\alpha$ is a weight factor; PixelLoss and LesionLoss are pixel- and lesion-level losses, respectively; $F_{focal}(\centerdot)$ is the focal loss, which can reduce the impact of category imbalance; and $k$ and $\gamma$ are constants.
The proposed compound loss function takes into account the recall loss and accuracy loss, which are weighted in LesionLoss and PixelLoss. The latter is defined as
\begin{equation}
\begin{aligned}
PixelLoss&=\beta \times Pre_{P}+(1-\beta)\times Rec_{P}\\
&=\beta \times (1-\frac{|T\cap P|+c}{|P|+c})\\
&\quad+(1-\beta)\times (1-\frac{|T\cap P|+c}{|T|+c})\vspace{1ex}
\end{aligned}
\label{eq}
\end{equation}
\noindent
where $\beta$ is a weighting factor, $Pre_{P}$ and $Rec_{P}$ represent pixel-level precision and recall loss, respectively.$|T\cap P|$ represents the number of pixels in the intersection of $T$ and $P$, meanwhile, $|\cap(T,P)|$ represents the number of lesions accurately predicted. LesionLoss is defined as
\begin{equation}
\begin{aligned}
LesionLoss&=\beta \times Pre_{L}+(1-\beta)\times Rec_{L}\\
&=\beta \times (1-\frac{|\cap(T,P)|+c}{|P|+c})\\
&\quad+(1-\beta)\times (1-\frac{|\cap(T,P)|+c}{|T|+c})\vspace{1ex}
\end{aligned}
\label{eq}
\end{equation}

The compound loss function combines pixel- and lesion-level loss, which ensures the improvement of the segmentation results of two levels. The calculation of lesion-level loss is based on the connected domain of a single lesion, where small and large lesions have the same weight.

By adding $\alpha$ and $\beta$, the compound loss function enables the tailoring of models according to use. A model trained with $\alpha<0.5$ has a better recall rate, and it provides more potential lesions for pathologists. When $\alpha=0.5$, the model accounts for the recall rate and accuracy and has more comprehensive performance. In practical applications, the intersection of the two models represents an area with a greater probability of hydrops lesions, and their difference represents an area with a certain probability of hydrops lesions.

\subsection{Stagewise Training Strategy}
For a variety of evaluation metrics, we propose a  stagewise training with the advantages of multiple loss functions, as shown in Fig. 5.
\begin{figure}[H]
	\centerline{\includegraphics[width=\columnwidth]{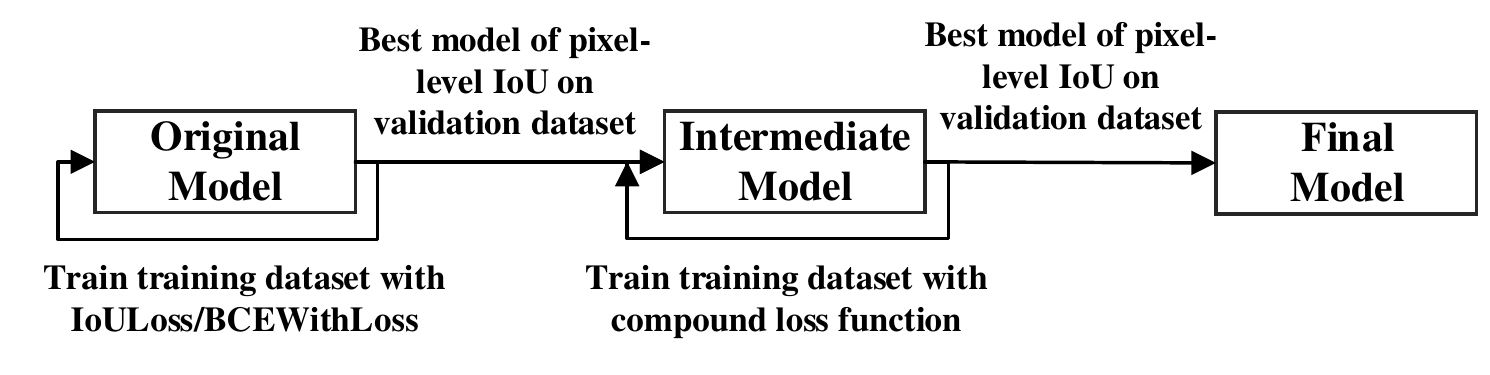}}
	\caption{Flow diagram of stagewise training strategy.}
	\label{fig}
\end{figure}

IoULoss or BCEWithLogitsLoss is used to train the model, and only the model with the best pixel-level IoU result on validation is saved, so as to ensure pixel-level performance. The model is trained on the compound loss function, and only the model with the best pixel-level IoU result on validation is saved, so as to maintain pixel-level performance and improve lesion-level performance. Experimental results show that this training method can effectively integrate the advantages of different loss functions and improve the performance of the model under multiple evaluation metrics.

\section{Experiment}

Forty-two sections with typical characteristics of hydrops lesions were selected from 157 sections as the section dataset.
\subsection{Preprocessing}
\subsubsection{Data cleaning}
The sliding method was used to crop HM sections to expand the dataset. The sliding window selects the 9000-pixel size according to the visual field size under the microscope. Imaging under this size basically ensures that the whole hydrops area of the villi can be seen in the cropped window. The sliding step is half the window size. Each cropped image (called “image” below) of the HM and the label mask constitute a set of training data. Since the lesion area in HM sections accounts for a small area, it is difficult to train a suitable hydrops lesion segmentation model when all  images are included in the dataset.

To reduce the class imbalance of the dataset, 3078 pieces of images with hydrops areas and 3724 pieces without hydrops areas were cleaned as training and validation dataset . As a result, 96 pieces of images without hydrops areas were randomly selected with probability 0.025. A total of 3174 pieces were split into train and valid at a ratio of 9:1. 

To evaluate the model performance on a new section, test images are from different subjects of training and validation dataset. The testing dataset finally obtained 330 images in the same cleaning way.
%% TABLE
\begin{table}[htbp]
	\caption{Image ratio of hydrops area and non-hydrops area}
	\label{table}
	\begin{tabular}{|l|l|p{1.4cm}|p{1.4cm}|}
		\hline
		\multicolumn{2}{|l|}{Dataset}                        & \begin{tabular}[c]{@{}l@{}}hydrops and \\ Non-\\hydrops \end{tabular} & \begin{tabular}[c]{@{}l@{}}hydrops and\\ 2.5\% Non-\\hydrops \end{tabular} \\ \hline
		\multirow{2}{*}{Train\&Valid} & hydrops area ratio     & 5.2\%                                                                    & 11.1\%                                                                        \\ \cline{2-4} 
		& Non-hydrops area ratio & 94.8\%                                                                   & 88.9\%                                                                        \\ \hline
		\multirow{2}{*}{Test}         & hydrops area ratio     & 4.3\%                                                                    & 15.6\%                                                                        \\ \cline{2-4} 
		& Non-hydrops area ratio & 95.7\%                                                                   & 84.4\%                                                                        \\ \hline
	\end{tabular}
\end{table}

The dataset retains part of the background so that the hydrops lesion segmentation model is more robust, with better discriminatory ability to prevent misdiagnosis. Data cleaning can increase the proportion of hydrops area to more than 10\% (see Table 1), which can prevent the impact of class imbalance.

\subsubsection{Data Augmentation}
Data enhancement expands the training set, with methods including horizontal reversal, rotation, scaling, and translation. Each method corresponds to a certain section situation. The final image transformation combines multiple image transformations, each with a probability of 0.5. Each batch performs random online data enhancement in each round of model training.

\subsubsection{Metrics}
There are pixel- and lesion-level metrics. A lesion-level evaluation indicator takes into account the needs of the pathological diagnosis of HM, as pathologists mainly consider a single lesion for diagnosis. The metrics are intersection over union (IoU), recall (Rec), and precision (Pre). IoU is for the comprehensive evaluation of lesion segmentation performance. Rec can evaluate the severity of a missed diagnosis, and Pre measures the amount of misdiagnosis.

\subsection{Segmentation Model Result}
\subsubsection{Segmentation network}
In the experiment, the input HM section image was $512 \times512\times3$ RGB. The Adam optimizer was used to train the network model. The initial learning rate was 0.0001, and the batch size was 8. All models used mean pooling, and the dropout parameter was set to 0.5. All the training data had random online data enhancement. Only the model with the best lesion segmentation performance was saved through training, so as to prevent overfitting. The saved model was used to predict the training, validation, and testing datasets to evaluate the performance of the hydrops lesion segmentation network.

We used DeepLabv3+, U-net, FPN, LinkNet, PSPNet, and pyramid attention network (PAN) [38] image segmentation networks. Since ResNet has excellent image feature-extraction ability and is commonly used in deep convolution networks, we used ResNet50 as the feature-extraction network.

Table 2 presents the parameters of different networks (M denotes million). The six networks have similar numbers of parameters. DeepLabv3+ and PSPNet are excellent in terms of time consumption for image processing. All the networks can basically meet the requirements of real-time detection.

\begin{table}[htbp]
	\centering
	\caption{Evaluation of different models}
	\label{table}
	\begin{tabular}{|p{1.8cm}|p{1.8cm}|p{1.8cm}|p{1.87cm}|}
		\hline
		Model   & Params(M) & Model Size(M) & Time(ms) \\ \hline
		DeepLabv3+ & 26.7  & 102.1     & 82.3     \\ \hline
		FPN        & 26.1  & 99.9      & 90.7     \\ \hline
		LinkNet    & 31.2  & 119.3     & 89.0     \\ \hline
		PAN        & 24.3  & 92.8      & 89.7     \\ \hline
		PSPNet     & 24.3  & 93.0      & 78.5     \\ \hline
		U-net       & 24.6  & 94.3      & 95.7     \\ \hline
	\end{tabular}
\end{table}

Fig. 6 and Table 3 show the performance of hydrops lesion segmentation in different networks on the test set. It can be seen that DeepLabv3+ performs better than the other networks.
\begin{figure*}[]
	\centerline{\includegraphics[width=19cm]{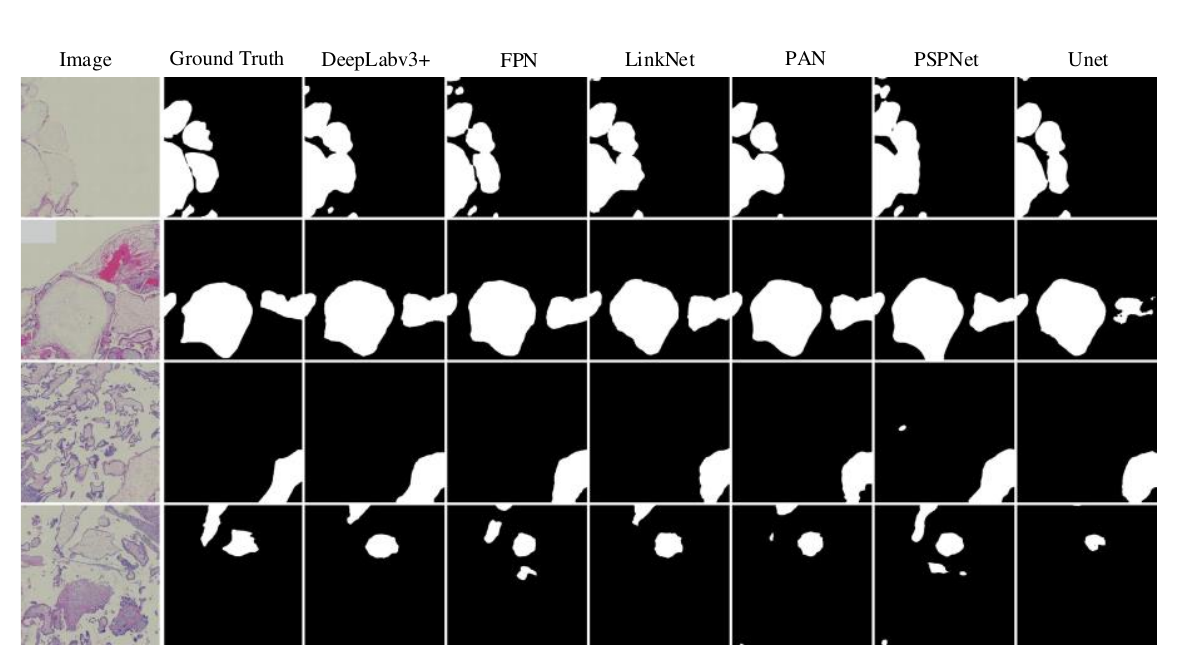}}
	\caption{Examples of hydrops lesions segmentation in different networks}
	\label{fig3}
\end{figure*}
In summary, considering time consumption, model size, hydrops lesion segmentation performance, and other factors, DeepLabv3+ is used as the main network in this paper.
\begin{table}[]
	\centering
	\caption{Evaluation performance of different Models on hydrops lesions segmentation}
	\label{table}
	\begin{tabular}{|l|p{0.7cm}|p{0.7cm}|p{0.7cm}|p{0.7cm}|p{0.7cm}|p{0.7cm}|}
		\hline
		\multirow{2}{*}{Model} & \multicolumn{3}{l|}{Pixel-level} & \multicolumn{3}{l|}{Lesion-level} \\ \cline{2-7} 
		& IoU(\%)   & Rec(\%)   & Pre(\%)  & IoU(\%)   & Rec(\%)   & Pre(\%)   \\ \hline
		DeepLabv3+                & 75.9      & 82.8      & 90.1     & 64.8      & 75.4      & 82.1      \\ \hline
		FPN                       & 74.5      & 85.5      & 85.2     & 61.8      & 76.6      & 76.2      \\ \hline
		LinkNet                   & 75.3      & 88.1      & 83.8     & 60.9      & 83.1      & 69.5      \\ \hline
		PAN                       & 67.1      & 72.5      & 90.0     & 64.0      & 76.8      & 79.3      \\ \hline
		PSPNet                    & 71.9      & 82.3      & 85.1     & 60.5      & 85.9      & 67.2      \\ \hline
		U-net                      & 75.6      & 90.7      & 82.0     & 61.7      & 85.8      & 68.8      \\ \hline
	\end{tabular}
\end{table}

\subsection{Compound Loss Function Analysis}
We conducted experiments on commonly used loss functions BCELoss, DiceLoss, IoULoss, and IoULoss-based FocalLoss, which outputs the model through a sigmoid function and calculates the loss between the model output and the real label.

The BCELoss can be calculated using
\begin{equation}
\begin{aligned}
BCEWithLogitsLoss
&=-\frac{1}{N}\sum_{i \in label}(y_i \times lnx_i\\ &+(1-y_i)\times ln(1-x_i)) \vspace{1ex}
\end{aligned}
\label{eq}
\end{equation}
\noindent
The DiceLoss can be calculated using 
\begin{equation}
\begin{aligned}
DiceLoss&=1-dice \\
&=1-\frac{2\times|T\times \sigma(P)|+c}{|T|+|\sigma(P)|+c} \vspace{1ex}
\end{aligned}
\label{eq}
\end{equation}
\noindent
The IoULoss can be calculated using
\begin{equation}
\begin{aligned}
IoULoss&=1-IoU \\
&=1-\frac{|T\times \sigma(P)|+c}{|T|+|\sigma(P)|-|T\times \sigma(P)|+c} \vspace{1ex}
\end{aligned}
\label{eq}
\end{equation}
\noindent

Where $N$ is the number of pixels in the label images, $y_i$ is the ground truth, $x_i$ is the prediction $T$ refers to the hydrops lesion area in ground truth and $P$ is the prediction hydrops lesion area, $|\centerdot|$is number of pixels, $\sigma(\centerdot)$is the sigmoid function, $c$ is a nonzero constant. 

In this experiment, DeepLabV3+ is used as the backbone network, and resnet50 is uniformly used for feature extraction network. The changing curves of IoU under different loss functions are shown in Fig.7. 
\begin{figure}[!t]
	\centerline{\includegraphics[width=9cm]{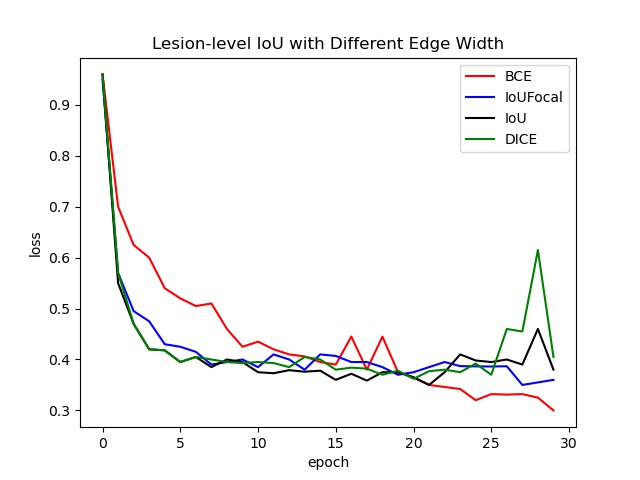}}
	\caption{Loss curve of  different loss functions.}
	\label{fig3}
\end{figure}

It can be seen that the curve of DiceLoss fluctuates in the last several rounds of training, and other loss functions tend to converge after 20 epoches. BCEWithLogitsLoss has not reached complete convergence, indicating that the model still has room for improvement.
Table 4 shows the evaluation results of hydrops lesions recognition with different loss functions on Test.

\begin{table}[]
	\centering
	
	\caption{Network performance with different Loss Function}
	\label{table}
	\begin{tabular}{|l|p{0.7cm}|p{0.7cm}|p{0.7cm}|p{0.7cm}|p{0.7cm}|p{0.7cm}|}
		
		\hline
		\multirow{2}{*}{$Loss Func.$} & \multicolumn{3}{l|}{Pixel-level} & \multicolumn{3}{l|}{Lesion-level} \\ \cline{2-7} 
		& IoU(\%)   & Rec(\%)   & Pre(\%)  & IoU(\%)   & Rec(\%)   & Pre(\%)   \\ \hline
		BCELoss                  & 75.9      & 82.4      & 90.7     & 66.4      & 75.1     & 85.2      \\ \hline
		DiceLoss                & 75.9      & 82.8      & 90.1     & 64.8      & 75.4      & 82.1      \\ \hline
		IoULoss                & 76.9      & 84.7      & 89.3     & 64.5      & 72.9      & 84.8      \\ \hline
		FocalLoss                & 73.6      & 79.4      & 90.1     & 62.1      & 68.3      & 87.3      \\ \hline
		
	\end{tabular}
\end{table}

We can see that BCELoss and IoULoss are slightly superior to other two loss functions. BCELoss is able to reach 66.4\% on lesion-level IoU and IoULoss is able to reach 76.9\% on pixel-level IoU. Each of the four loss functions has its own advantages and disadvantages, but no single loss function is able to achieve satisfactory results in both pixel- and lesion-level evaluation indices.

Moving on to our proposed compound loss function, we tested different weight coefficients $\alpha$ in the compound loss function, with results as shown in Table 5.

\begin{table}[htbp]
	\centering
	\caption{Network performance with different $\alpha$}
	\label{table}
	\begin{tabular}{|l|l|l|l|l|l|l|}
		\hline
		\multirow{2}{*}{$\alpha$} & \multicolumn{3}{l|}{Pixel-level} & \multicolumn{3}{l|}{Lesion-level} \\ \cline{2-7} 
		& IoU(\%)   & Rec(\%)   & Pre(\%)  & IoU(\%)   & Rec(\%)   & Pre(\%)   \\ \hline
		0                  & 72.4      & 80.4      & 87.9     & 60.4      & 81.8      & 69.7      \\ \hline
		0.2                & 74.6      & 85.1      & 85.8     & 62.4      & 80.7      & 73.3      \\ \hline
		0.4                & 74.7      & 88.2      & 83.0     & 62.4      & 85.2      & 70.1      \\ \hline
		0.5                & 76.1      & 84.8      & 88.1     & 67.1      & 79.0      & 81.6      \\ \hline
		0.6                & 74.2      & 84.1      & 86.3     & 63.8      & 77.0      & 78.7      \\ \hline
		0.8                & 75.1      & 84.4      & 87.1     & 61.9      & 76.3      & 76.6      \\ \hline
		1.0                & 19.5      & 96.7      & 19.6     & 81.1      & 100.0     & 81.1      \\ \hline
	\end{tabular}
\end{table}

With the increase of lesion-level loss weight $\alpha$, lesion- and pixel-level IoU are significantly improved, indicating that adding lesion-level loss to the loss function can improve lesion-level performance and bring positive benefits at the pixel level. The lesion- and pixel-level IoU are maximized when $\alpha=0.5$. When $\alpha>0.5$ and keeps increasing, both these decrease. Hence, under the premise of ensuring that the pixel-level loss fully affects the segmentation accuracy of each pixel in the network, the appropriate lesion-level loss is added to improve the segmentation accuracy of the model. When $\alpha=1$, the pixel-level IoU drops sharply, whereas the lesion-level IoU rises sharply. When the loss function is composed of lesion-level losses, the prediction label tends to be a whole white map, which illustrates the necessity of pixel-level loss in the loss function.

\begin{table}[]
	\centering
	\caption{Network performance with different $\beta$}
	\label{table}
	\begin{tabular}{|l|l|l|l|l|l|l|}
		\hline
		\multirow{2}{*}{$\beta$} & \multicolumn{3}{l|}{Pixel-level} & \multicolumn{3}{l|}{Lesion-level} \\ \cline{2-7} 
		& IoU(\%)   & Rec(\%)   & Pre(\%)  & IoU(\%)   & Rec(\%)   & Pre(\%)   \\ \hline
		0                  & 40.5      & 99.8      & 40.5     & 48.4      & 100.0     & 48.4      \\ \hline
		0.1                & 69.0      & 96.3      & 70.8     & 56.3      & 96.3      & 57.5      \\ \hline
		0.3                & 72.8      & 91.2      & 78.3     & 56.1      & 89.0      & 60.3      \\ \hline
		0.5                & 76.1      & 84.8      & 88.1     & 67.1      & 79.0      & 81.6      \\ \hline
		0.6                & 72.1      & 78.6      & 89.8     & 61.0      & 72.9      & 78.8      \\ \hline
		0.7                & 60.3      & 62.4      & 94.8     & 52.3      & 58.1      & 83.9      \\ \hline
		0.9                & 43.2      & 45.7      & 88.5     & 47.0      & 54.0      & 78.5      \\ \hline
	\end{tabular}
\end{table}

A similar experiment evaluated different weight coefficients $\beta$. Table 6 shows that when $\beta=0.5$, the lesion- and pixel-level IoU are maximized, which means that IoU must equally consider Pre and Rec losses. Moreover, as $\beta$ deviates from 0.5, IoU obviously decreases. With the increase of $\beta$, the lesion area becomes smaller, the accuracy rate of the lesion area becomes larger, and the recall rate becomes smaller. In actual pathological diagnosis, the requirement of recall rate is stricter, which means a missed diagnosis is more severe than a misdiagnosis. Therefore, the model with $\beta=0.3$ can be selected as a standby, which is suitable for pathologists who demand a high recall rate.
\begin{figure*}[]
	\centerline{\includegraphics[width=17cm]{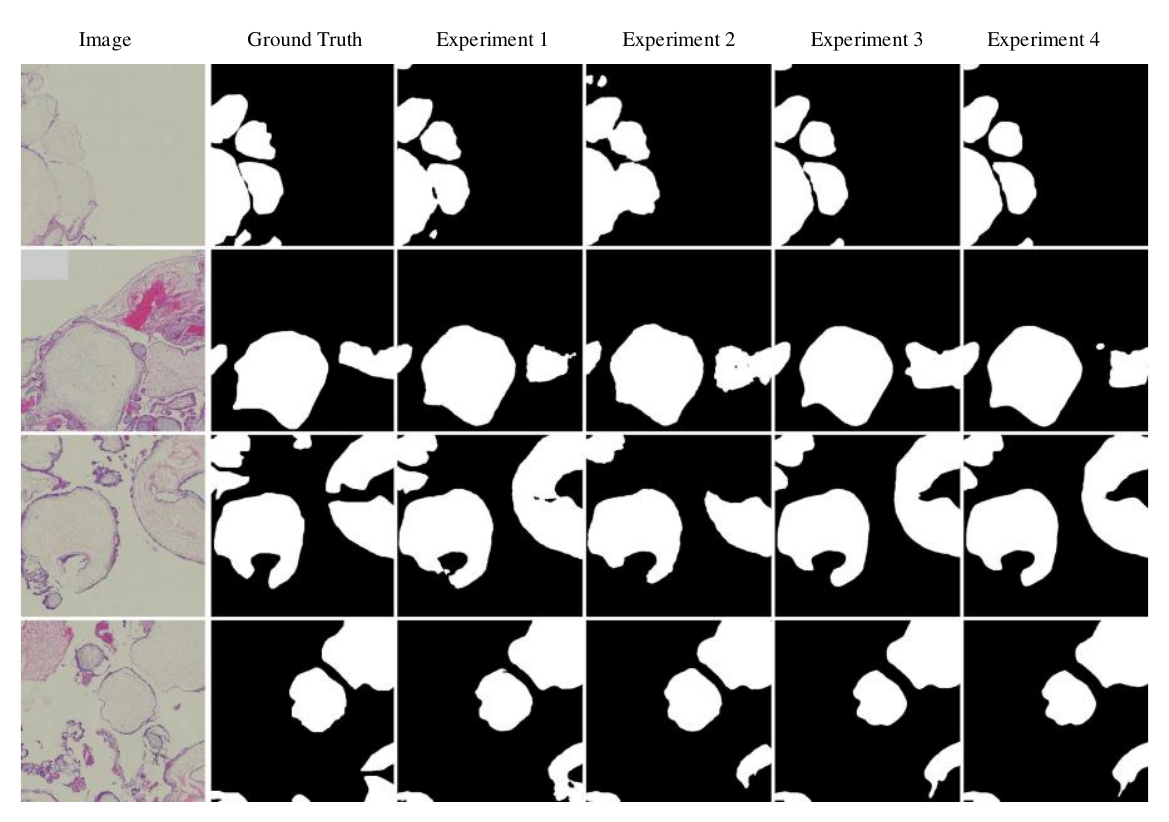}}
	\caption{Examples of hydrops lesions segmentation in different stagewise training method}
	\label{fig3}
\end{figure*}
\subsection{Stagewise Training Method Result}
  We performed four experiments to verify the effect of stagewise training based on multiple loss functions:

a) Experiment 1: learning rate is 1e-4, and IoULoss is used to train the model for 50 epochs. The learning rate is changed to 1e-5, and the compound loss function with $\alpha=0.5$ and $\beta=0.5$ is used for another 50 epochs;

b) Experiment 2: learning rate is 1e-4, and IoULoss is used to train the model for 50 epochs. The learning rate is changed to 1e-5, and IoULoss is used for another 50 epochs;

c) Experiment 3: learning rate is 1e-4, and BCEWithLogitsLoss is used to train the model for 50 epochs. The learning rate is changed to 1e-5, and the compound loss function with $\alpha=0.5$ and $\beta=0.5$ is used for another 50 epochs;

d) Experiment 4: learning rate is 1e-4, and BCEWithLogitsLoss is used to train the model for 50 epochs. The learning rate is changed to 1e-5, and BCEWithLogitsLoss is used for another 50 epochs.
The results using DeepLabv3+ are shown in Table 7 and Fig. 8.
\begin{table}[H]
	
	\caption{Performance evaluation of multi-loss function model by stagewise training method}
	\label{table}
	\begin{tabular}{|l|l|l|l|l|l|l|}
		\hline
		\multirow{2}{*}{\begin{tabular}[c]{@{}l@{}}Experi-\\ ment\end{tabular}} & \multicolumn{3}{l|}{Pixel-level} & \multicolumn{3}{l|}{Lesion-level} \\ \cline{2-7} 
		& IoU(\%)   & Rec(\%)   & Pre(\%)  & IoU(\%)   & Rec(\%)   & Pre(\%)   \\ \hline
		E1                                                                      & 77.0      & 88.1      & 86.0     & 70.2      & 86.2      & 79.1      \\ \hline
		E2                                                                      & 76.1      & 88.6      & 84.3     & 68.2      & 79.6      & 82.6      \\ \hline
		E3                                                                      & 76.1      & 87.9      & 85.0     & 67.4      & 81.5      & 79.6      \\ \hline
		E4                                                                      & 75.4      & 87.9      & 84.2     & 66.4      & 80.4      & 79.2      \\ \hline
	\end{tabular}
\end{table}

From the comparison of Experiment 2(called E2 below) and E4, it can be seen that IoULoss is superior to BCEWithLogitsLoss on the HM hydrops dataset. From E1-E2 group and E3-E4 group, after adding the compound loss function, lesion-level IoU and recall has significant advance, the pixel-level IoU and precision have also increased. This shows that the stagewise training of the model based on multiple loss functions can further improve the model on multiple metrics.

\section{Conclusion}
We constructed an HM hydrops lesion detection model based on a semantic segmentation network to improve the efficiency of HM diagnosis and reduce misdiagnosis.

We completed the HM hydrops lesion dataset collection and annotation and established the dataset for the hydrops lesion segmentation model by using a sliding window to crop sections into image patches, with data cleaning and online data enhancement. Models under different networks, feature-extraction networks, and loss functions were tested and evaluated, with DeepLabv3+ as the segmentation network and se\_resnet50 as the feature-extraction module to segment lesions. Most importantly, we proposed a compound loss function for multiple evaluation metrics, which was proved to be superior to the traditional loss function in multiple evaluation indexes through comparative experiments. The performance of the model was significantly improved with the proposed stagewise training method of multiple loss functions.

Since the labeled dataset included hydrops, hyperplasia, and villi, model training was conducted on the hydrops dataset. In the future, it is hoped that models can be trained on the hyperplasia dataset, which can display hydrops and hyperplasia lesions in real time, to assist pathologists in the diagnosis of HM slices. 

The uses of our method are not limited to assisting HM diagnosis. Despite the morphological diversity of the tissues and organs, the basic ideas for pathological diagnosing are similar, to find the corresponding pathological manifestations, the lesion. Most lesion have different pathological features comparing to normal tissues and organs. Extracting those pathological features can make a difference in the diagnosing process and workflow. And our method has shown its powerful feature extraction capability in the paper. Besides, all of the components of our method are non-specified. Therefore, the model we proposed in this paper is not only to help with HM diagnosis but also has the ability to provide help with other pathologic diagnoses in the form of extracting the corresponding pathological features.


 \section*{Acknowledgment}
 The authors would like to thank \emph{the Third Affiliated Hospital of Zhengzhou University} for providing the data source.


\begin{thebibliography}{00}
 	
 	\bibitem{b1} H. O. Smith, ``Gestational Trophoblastic Disease Epidemiology and Trends,'' \emph{Clinical Obstetrics and Gynecology}, vol. 46, no. 3, pp. 541-556, Sep. 2003.
 	
 	\bibitem{b2} N. Sebire, R. Fisher and H. Rees, ``Histopathological Diagnosis of Partial and Complete Hydatidiform Mole in the First Trimester of Pregnancy,'' \emph{Pediatr. Dev. Pathol.}, vol. 6, no. 1, pp. 69-77, Jan. 2003.
 	
 	\bibitem{b3} P. Palee, B. Sharp, L. Noriega, N. Sebire and C. Platt, ``Heuristic neural network approach in histological sections detection of hydatidiform mole,'' \emph{J. Med. Imag.}, vol. 6, no. 4, pp. 1-9, Nov. 2019.
 	
 	\bibitem{b4} C.-Y. Zeng, Y.-B. Chen, L.-J. Zhao and B. Wan, ``Partial hydatidiform mole and coexistent live fetus: a case report and review of the literature,'' \emph{Open Medicine}, vol. 14, no. 1, pp. 843-846, Nov. 2019.
 	
 	\bibitem{b5}  J.-J. Candelier, ``The hydatidiform mole,'' \emph{Cell Adhesion \& Migration}, vol. 10, no. 1-2, pp. 226-235, Mar. 2016.
 	
 	\bibitem{b6}  N. M. Lindor, J. A. Ney, T. A. Gaffey, R. B. Jenkins, S. N. Thibodeau, and G. W. Dewald, ``A Genetic Review of Complete and Partial Hydatidiform Moles and Nonmolar Triploidy,'' \emph{Mayo Clinic Proceedings}, vol. 67, no. 8, pp. 791-799, Aug. 1992.
 	
 	\bibitem{b7}   G. Mangili et al, ``Hydatidiform mole: age-related clinical presentation and high rate of severe complications in older women,'' \emph{Acta Obstet Gynecol Scand}, vol. 93, no. 5, pp. 503-507, May. 2014.
 	
 	\bibitem{b8}   L.-Z. Jiao, S.-Y. You, Y.-P. Wang, C.-G. Zhu and J.-Y. Jiang, ``Clinical characteristics and diagnosis of early hydatidiform mole,'' \emph{Chung-hua fu ch'an k'o tsa chih}, vol. 54, no. 11, pp. 756-762, Nov. 2019.
 	
 	\bibitem{b9}   R. Nock and F. Nielsen, ``Statistical region merging,'' \emph{IEEE Transactions on pattern analysis and machine intelligence}, vol. 26, no. 11, pp. 1452-1458, 2004.
 	
 	\bibitem{b10}   N. Dhanachandra, K. Manglem, and Y. J. Chanu, ``Image segmentation using k-means clustering algorithm and subtractive clustering algorithm,'' \emph{Procedia Computer Science}, vol. 54, pp. 764-771, 2015.
 	
 	\bibitem{b11}   Y. Boykov, O. Veksler, and R. Zabih, ``Fast approximate energy minimization via graph cuts,'' \emph{IEEE Transactions on pattern analysis and machine intelligence}, vol. 23, no. 11, pp. 1222-1239, 2001.
 	
 	\bibitem{b12}   J.-L. Starck, M. Elad, and D. L. Donoho, ``Image decomposition via the combination of sparse representations and a variational approach,'' \emph{IEEE transactions on image processing}, vol. 14, no. 10, pp. 1570-1582, 2005.
 	
 	\bibitem{b13}   S. Minaee, Y. Y. Boykov, F. Porikli, A. J. Plaza, N. Kehtarnavaz and D. Terzopoulos, ``Image Segmentation Using Deep Learning: A Survey,'' \emph{ IEEE Transactions on Pattern Analysis and Machine Intelligence}, pp. 1-1, 2021, DOI: 10.1109/TPAMI.2021.3059968.
 	
 	\bibitem{b14}   J. Long, E. Shelhamer, and T. Darrell, ``Fully convolutional networks for semantic segmentation,'' 2015, pp. 3431-3440.Accessed: Jan. 06, 2022. [Online].
 	Available: \underline{https://openaccess.thecvf.com}
 	
 	\bibitem{b15}   O. Ronneberger, P. Fischer, and T. Brox, ``U-net: Convolutional networks for biomedical image segmentation,'' in \emph{Medical Image Computing and Computer-Assisted Intervention – MICCAI 2015} Cham, 2015, pp. 234-241. DOI: 10.1007/978-3-319-24574-4\_28. 
 	
 	\bibitem{b16}   T.-Y. Lin, P. Dollar, R. Girshick, K. He, B. Hariharan, and ´S. Belongie, ``Feature pyramid networks for object detection,'' in \emph{2017 IEEE Conference on Computer Vision and Pattern Recognition (CVPR)} Honolulu, HI, 2017,   pp. 936-944.
 	
 	\bibitem{b17}   V. Badrinarayanan, A. Kendall, and R. Cipolla, ``Segnet: A deep convolutional encoder-decoder architecture for image segmentation,'' \emph{IEEE transactions on pattern analysis and machine intelligence}, vol. 39, no. 12, pp. 2481-2495, 2017.
 	
 	\bibitem{b18}    H. Zhao, J. Shi, X. Qi, X. Wang, and J. Jia, ``Pyramid scene parsing network,'' 2017, pp. 2881-2890.Accessed: Jan. 06, 2022. [Online].
 	Available: \underline{https://openaccess.thecvf.com}
 	
 	\bibitem{b19}   A. Chaurasia and E. Culurciello, ``Linknet: Exploiting encoder representations for efficient semantic segmentation,'' in \emph{2017 IEEE Visual Communications and Image Processing (VCIP)}, Dec. 2017, pp. 1-4. DOI: 10.1109/VCIP.2017.8305148.
 	
 	\bibitem{b20}   L.-C. Chen, G. Papandreou, I. Kokkinos, K. Murphy, and A. L. Yuille, ``Deeplab: Semantic image segmentation with deep convolutional nets, atrous convolution, and fully connected crfs,'' \emph{IEEE transactions on pattern analysis and machine intelligence}, vol. 40, no. 4, pp. 834-848, 2017.
 	
 	\bibitem{b21}    L.-C. Chen, G. Papandreou, F. Schroff, and H. Adam, ``Rethinking Atrous Convolution for Semantic Image Segmentation,'',Dec. 2017, Accessed: Jan. 06, 2022. [Online].
 	Available: \underline{http://arxiv.org/abs/1706.05587}
 	
 	\bibitem{b22}    L.-C. Chen, Y. Zhu, G. Papandreou, F. Schroff, and H. Adam, ``Encoder-Decoder with Atrous Separable Convolution for Semantic Image Segmentation,'' 2018, pp. 801-818.Accessed: Jan. 06, 2022. [Online].
 	Available: \underline{https://openaccess.thecvf.com}
 	
 	\bibitem{b23}  K. M. Rajpoot and N. M. Rajpoot, ``Wavelet based segmentation of hyperspectral colon tissue imagery,'' in \emph{7th International Multi Topic Conference}, 2003, pp. 38-43. DOI: 10.1109/INMIC.2003.1416612.
 	
 	\bibitem{b24}   N. Kumar et al., ``Hyperspectral Tissue Image Segmentation Using Semi-Supervised NMF and Hierarchical Clustering,'' \emph{IEEE Transactions on Medical Imaging}, vol. 38, no. 5, pp. 1304-1313, May. 2019.
 	
 	\bibitem{b25}   A. C. Simsek, A. B. Tosun, C. Aykanat, C. Sokmensuer, and C. Gunduz-Demir, ``Multilevel Segmentation of Histopathological Images Using Cooccurrence of Tissue Objects,'' \emph{IEEE Transactions on Biomedical Engineering}, vol. 59, no. 6, pp. 1681-1690, Jun. 2012.
 	
 	\bibitem{b26}   L. Gorelick et al., ``Prostate Histopathology: Learning Tissue Component Histograms for Cancer Detection and Classification,'' \emph{IEEE Transactions on Medical Imaging}, vol. 32, no. 10, pp. 1804-1818, Oct. 2013.
 	
 	\bibitem{b27}   P. Palee, B. Sharp, L. Noriega, N. J. Sebire, and C. Platt, ``Image analysis of histological features in molar pregnancies,'' \emph{Expert Systems with Applications}, vol. 40, no. 17, pp. 7151-7158, Dec. 2013.
 	
 	\bibitem{b28}   Y. Zheng et al., ``Diagnostic Regions Attention Network (DRA-Net) for Histopathology WSI Recommendation and Retrieval,'' \emph{IEEE Transactions on Medical Imaging}, vol. 40, no. 3, pp. 1090-1103, Mar. 2021.
 	
 	\bibitem{b29}   Y. Dong et al., ``A Polarization-Imaging-Based Machine Learning Framework for Quantitative Pathological Diagnosis of Cervical Precancerous Lesions,'' \emph{IEEE Transactions on Medical Imaging}, vol. 40, no. 12, pp. 3728-3738, Dec. 2021.
 	
 	\bibitem{b30}   Q.-K. Liang et al., ``Feasibility of deep learning image-based segmentation algorithm in pathological section of gastric cancer,'' \emph{Academic Journal of Second Military Medical University}, vol. 39, no. 08, pp. 903-908, 2018.
 	
 	\bibitem{b31}  X. Wang, G. Yu, Z. Yan, L. Wan, W. Wang, and L. C. Cui Lizhen, ``Lung Cancer Subtype Diagnosis by Fusing Image-genomics Data and Hybrid Deep Networks,`` \emph{IEEE/ACM Transactions on Computational Biology and Bioinformatics}, pp. 1–1, 2021, doi: 10.1109/TCBB.2021.3132292.
 	
 	
 	\bibitem{b32}   T. Wan, J. Cao, J. Chen, and Z. Qin, ``Automated grading of breast cancer histopathology using cascaded ensemble with combination of multi-level image features,'' \emph{Neurocomputing}, vol. 229, pp. 34-44, Mar. 2017.
 	
 	\bibitem{b33}   Y. Li, ``CRU-Net: A Deep Learning Network for Semantic Segmentation of Pathological Tissue Slices,'' in \emph{2021 IEEE International Conference on Artificial Intelligence and Industrial Design (AIID)}, May 2021, pp. 46-50. DOI: 10.1109/AIID51893.2021.9456469.
 	
 	\bibitem{b34}  Z. Meng, Z. Zhao, B. Li, F. Su, L. Guo, and H. Wang, ``Triple Up-Sampling Segmentation Network With Distribution Consistency Loss for Pathological Diagnosis of Cervical Precancerous Lesions,'' \emph{IEEE Journal of Biomedical and Health Informatics}, vol. 25, no. 7, pp. 2673-2685, Jul. 2021.
 	
 	\bibitem{b35}   Z. Zhang et al., ``Pathologist-level interpretable whole-slide cancer diagnosis with deep learning,'' \emph{Nat Mach Intell}, vol. 1, no. 5, pp. 236-245, May. 2019.
 	
 	
 	
 	\bibitem{b36} J. Feng, S. Zhang, and L. Chen, ''Extracting ROI-Based Contourlet Subband Energy Feature from the sMRI Image for Alzheimer’s Disease Classification'', \emph{/ACM Transactions on Computational Biology and Bioinformatics}, pp. 1–1, 2021, doi: 10.1109/TCBB.2021.3051177.
 	
 	\bibitem{b37} J. Cheng et al., ‘Automated Diagnosis of COVID-19 using Deep Supervised Autoencoder with Multi-view Features from CT Images’, \emph{IEEE/ACM Transactions on Computational Biology and Bioinformatics}, pp. 1–1, 2021, doi: 10.1109/TCBB.2021.3102584.
 	
	\bibitem{b38} H. Li, P. Xiong, J. An, and L. Wang, “Pyramid Attention Network for Semantic Segmentation,” arXiv:1805.10180 [cs], May 2018, Accessed: Feb. 25, 2022. [Online]. Available: http://arxiv.org/abs/1805.10180
 	
 	
 	
 \end{thebibliography}
\end{document}